\documentclass[fleqn,twoside]{article}      % Specifies the document class
\usepackage{epsf,multicol,ifthen}
\usepackage{ujp}
\usepackage[cp1251]{inputenc}
\usepackage[english,ukrainian]{babel}
\usepackage{amstext}
\usepackage{amssymb}
\usepackage{cite}
\usepackage {graphicx}

\nazva{ON CHEBYSHEV POLYNOMIALS AND TORUS KNOTS}%
\udk{02.10.Kn, 02.20.Uw}
\nazvacol{ON CHEBYSHEV POLYNOMIALS AND TORUS KNOTS}%
\avtor{A.M. GAVRILIK, A.M. PAVLYUK}%
\avtorcol{A.M. GAVRILIK, A.M. PAVLYUK}%
\inst{Bogolyubov Institute for Theoretical Physics, Nat. Acad. Sci. of Ukraine} %
\adr{(14b, Metrolohichna Str., Kyiv 03680, Ukraine; e-mail: omgavr@bitp.kiev.ua;
                                                            pavlyuk@bitp.kiev.ua)} %

\newcommand{\ed}{\end{document}}
\newcommand{\be}{\begin{equation}}
\newcommand{\ee}{\end{equation}}
\newcommand{\bc}{\begin{center}}
\newcommand{\ec}{\end{center}}
\newcommand{\ba}{\begin{array}}
\newcommand{\ea}{\end{array}}

\newcommand{\wt}{\widetilde}

\begin{document}
\setcounter{page}{129}%
\maketitl                 % Produces the title.
\begin{multicols}{2}
\anot{%
In this work we demonstrate that the $q$-numbers and their
two-parameter generalization, the $q,\!p$\,-numbers, can be used to
obtain some polynomial invariants for torus knots and links.
 First, we show that the $q$-numbers, which are closely connected with the
Chebyshev polynomials, can also be related with the Alexander
polynomials for the class $T(s,2)\,$ of torus knots, $s$ being an
odd integer, and used for finding the corresponding skein relation.
Then, we develop this procedure in order to obtain, with the help of
$q,p$\,-numbers, the generalized two-variable Alexander polynomials,
and prove their direct connection with the
HOMFLY polynomials and the skein relation of the latter.}%

\section{Introduction}

The relevance of knots and links to many physical~\cite{At, Ka1,
Ka2} and biophysical~\cite{FV} systems implies the importance of
investigating the properties and characteristics of knot-like
structures. The concepts of knot theory play important role in the
models of statistical physics~\cite{AW}, quantum field
theory~\cite{Wi}, quantum gravity~\cite{Ba} and in a number of other
physical phenomena. In the preprint of 1975 it was proposed by L.D.
Faddeev that knot-like solitons could be realized in a nonlinear
field theory~\cite{Fa}, in a definite model defined in $3+1$
dimensions. The model includes the standard nonlinear $O(3)$
$\sigma$-model, which admits static solitons in $2+1$ dimensions,
and a Skyrme term. In the Faddeev model static solitons are
stabilized by the integer-valued Hopf charge. Interest in the model
was renewed in 1997 after an article of Faddeev and Niemi in
Nature~\cite{FN}. They have made first attempts at a numerical
construction of  solitons with the minimal energy in the form of
knots. Battye and Sutcliffe demonstrated that for higher Hopf charge
twisted, knotted and linked configurations occur~\cite{BS-98}, in
particular, they showed that the minimal energy soliton with Hopf
charge seven is a trefoil knot.

 R.J. Finkelstein has proposed  a field theory model, in which
 local $\, SU(2)\times
U(1)\,,$ the symmetry group of the standard electroweak theory, is
combined with global quantum group $SU_{q}(2)\,,$ the symmetry group
of knotted solitons~\cite{Fi1, Fi2}. This  allows to incorporate the
$q$-soliton into the field theory and to replace the point particles
by knotted solitons. More recent discussion on the role of field
theory knots both in superconductivity theory and in the Yang-Mills
theory can be found in~\cite{Fa2}.

In the context of modelling static properties of hadrons, it was
shown in~\cite{Ga1} (see also~\cite{Ga2}) that global quantum groups
$SU_{q}(n)\,,\,n=2,\ldots ,6\,,$ can be successfully applied for
flavor symmetries, and certain torus knots put into correspondence,
through Alexander polynomials, with vector quarkonia.

Various polynomial invariants are known to be one of the basic
characteristics of knots and links (see e.g.~\cite{Bir}). Among them,
the Alexander polynomials, the Jones polynomials and the HOMFLY
polynomials are both best studied and play important role in the
knot theory and its applications.

To describe some properties and characteristics of knots and links
the classical Chebyshev polynomials can be used. For example,
in~\cite{KP} the Chebyshev polynomials were utilized for polynomial
parametrization of noncompact counterparts of torus knots. It was
shown how to construct the Chebyshev model associated with any knot.

In this paper we concentrate on studying the polynomial invariants
such as the Alexander polynomials and the HOMFLY polynomials,
%one of the main characteristics of knots and links,
and their close connection with the Chebyshev polynomials. We
restrict ourselves with the set of torus knots and links, and show
that certain rather simple two-variable generalization of the
Chebyshev polynomials is well suited for characterizing those knots.

\section{Alexander polynomials and skein relation }

 The Alexander polynomials $A(t)$ for knots and links can be defined
 (see e.g.~\cite{Bir})
by the skein relation
 \be\label{alex-skein}
A_{+}(t)=(t^{1\over2}-t^{-{1\over2}})A_{O}(t)+A_{-}(t)\,,
 \ee
 and the condition for the unknot: \be\label{unknot} A_{unknot}=1\,.\ee
Using (\ref{alex-skein}) and (\ref{unknot}), one can find the
Alexander polynomial for any knot or link applying to it, in a
standard way, the surgery operations of switching and elimination.

From now on we consider the torus knots and links of type $(s,2)\,,$
where $s$ is any positive integer.
 If $s=1\,$ we have the unknot, the case of $s=2\,$ corresponds to the Hopf
 link, and $s=3\,$ to the trefoil knot, and so on. In general, when $s$
is odd, we have the series of torus knots $T(s,2)\,,$ and if $s$ is
even, we have the series of two-component torus links $L(s,2)\,.$
Here $s$ equals the minimal number of crossings.

Applying the operation of elimination to $(s,2)\,,$ one obtains
$(s-1,2)\,,$ whereas the switching operation turns $(s,2)\,$ into
$(s-2,2)\,,$ for $s>2\,.$ This means that $A_{+}(t),\, A_{O}(t)$ and
$A_{-}(t)$ correspond to three successive Alexander polynomials,
%describing knots and links from the set $(s,2),\,\, s=1,2,3\ldots\,, $
which allows to make the following
juxtaposition in (\ref{alex-skein})
%(where $n+1$ stands for $s$)
\be\ba{l} \label{cor}A_{+}(t)\rightarrow{\wt A}_{n+1}^{2}(t)\,,\quad
A_{O}(t) \rightarrow{\wt A}_{n}^{2}(t)\,,
\vspace{2mm}\\
A_{-}(t)\rightarrow{\wt A}_{n-1} ^{2}(t)\,. \ea\ee
 Thus, from
(\ref{alex-skein}) and (\ref{cor}) one obtains the recurrence
relation for the tilded Alexander polynomials for the unified set of
torus knots and links of type $(s,2)\,,$ (the polynomials are
arranged by increasing degrees):
 \be\label{alex-rec} {\wt
A}_{n+1}^{2}(t)=(t^{1\over2}-t^{-{1\over2}}){\wt A}_{n}^{2}(t)+{\wt
A}_{n-1}^{2}(t)\,. \ee

It is convenient to denote the Alexander polynomials for the subset
containing only torus knots (or the subset of torus links) as
%$$A(s,2)(t)\equiv A(2m+1,2)(t)\equiv
\[ A_{m}^{s,2}(t)\equiv A_{m}^{s,2}\equiv A_{m}^{2} \,.\]
Here $m$ is the  degree of the corresponding Alexander polynomial,
which has the form of Laurent polynomial:
\be\label{max}m={1\over2}(s-1)\,.\ee
  Since $s=2m+1\,$ for both knots and links,  the degree
$m$ of the Alexander polynomial for knots $T(s,2)\,$ is an integer,
and $m$ for links $L(s,2)\,$ is half-integer.
 Let us first give the table of the
Alexander polynomials $\,A_{m}^{s,2}(t)$  for torus knots
$\,T(s,2)\equiv T(2m+1,2)\,$ \be \label{tor2} \ba{l}
%A(1,2)(q)\equiv
A_{0}^{1,2}(t)=1\,,
\vspace{2mm}\\
%A(3,2)(q)\equiv
A_{1}^{3,2}(t)=t-1+t^{-1}\,,
\vspace{2mm}\\
%A(5,2)(q)\equiv
A_{2}^{5,2}(t)=t^{2}-t+1-t^{-1}+t^{-2}\,,
\vspace{2mm}\\
%A(7,2)(q)\equiv
A_{3}^{7,2}(t)=t^{3}-t^{2}+t-1+t^{-1}-t^{-2}+t^{-3}\,,
\vspace{2mm}\\
%A(9,2)(q)\equiv
%A_{4}^{9,2}(t)=t^{4}-t^{3}+t^{2}-t+1-t^{-1}+t^{-2}-t^{-3}+t^{-4}\,,
%\vspace{2mm}\\
\cdots\cdots\cdots\cdots\cdots\cdots\cdots\cdots\cdots\cdots\cdots\cdots\cdots\cdots
%\cdots\cdots\cdots\cdots\cdots\cdots
\vspace{2mm}\\
%A(2m+1,2)(t)\equiv
A_{m}^{2m+1,2}(t)=t^{m}-t^{m-1}+
%t^{m-2}-
\cdots -t^{-(m-1)}+t^{-m}=
\vspace{2mm}\\
=\sum\limits_{i=0}^{m}t^{m-2i}-\sum\limits_{i=0}^{m-1}t^{m-2i-1}=
\sum\limits_{i=0}^{2m}(-1)^{i}t^{m-i}\,.
\end{array}\end{equation}
Recurrence formula for the polynomials (\ref{tor2}) looks as
(dropping $2m+1$ in superscript)
 \be\label{alex-rec2}
A_{m+1}^{2}(t)=(t+t^{-1})A_{m}^{2}(t)-A_{m-1}^{2}(t)\,. \ee

Now consider torus links of the type $\,L(s,2)\equiv L(2m+1,2)\,,$
where $s$ is positive even integer. The degree $m$ of the Alexander
polynomial $\,A_{m}^{s,2}(q)\,$ is again as in (\ref{max}). It is
half-integer now.
The table shows the Alexander polynomials for these torus links: % ($m$ is half integer):
\be
\label{tor2-link}
\ba{l}
%A(2,2)(t)\equiv
A_{1\over2}^{2,2}(t)=t^{1\over2}-t^{-{1\over2}},
\vspace{2mm}\\
%A(4,2)(t)\equiv
A_{3\over2}^{4,2}(t)=t^{3\over2}-t^{1\over2}+t^{-{1\over2}}-t^{-{3\over2}}\,,
\vspace{2mm}\\
%A(6,2)(t)\equiv
A_{5\over2}^{6,2}(t)=t^{5\over2}-t^{3\over2}+t^{1\over2}-t^{-{1\over2}}+
t^{-{3\over2}}-t^{-{5\over2}}\,,
\vspace{2mm}\\
%A(8,2)(t)\equiv
%A_{7\over2}^{8,2}(t)=t^{7\over2}-t^{5\over2}+t^{3\over2}-t^{1\over2}+t^{-{1\over2}}
%-t^{-{3\over2}}+t^{-{5\over2}}-t^{-{7\over2}}\,,
%\vspace{2mm}\\
%A(9,2)(q)\equiv A_{4}^{9,2}(q)=q^{4}-q^{3}+q^{2}-q+1-q^{-1}+q^{-2}-q^{-3}+q^{-4}\,,
%\vspace{2mm}\\
\cdots\cdots\cdots\cdots\cdots\cdots\cdots\cdots\cdots\cdots\cdots\cdots\cdots\cdots
%\cdots\cdots\cdots\cdots\cdots\cdots
\vspace{2mm}\\
%A(2m+1,2)(t)\equiv
A_{m}^{2m+1,2}(t)=t^{m}-t^{m-1}+
%t^{m-2}-
\cdots +t^{-(m-1)}-t^{-m}=
\vspace{2mm}\\
=\sum\limits_{i=0}^{2m}(-1)^{i}t^{m-i}\,. \ea \ee
 Note, the
polynomials (\ref{tor2-link}) satisfy the recurrence relation
(\ref{alex-rec2}) too, though now with half-integer subscripts.
 So, for torus knots and links of type $(s,2)\,,$ eq. (\ref{alex-rec2})
describes either three successive torus knots (if $s$ is an odd
integer), or three successive torus links (if $s$ is an even
integer). Unifying two tables (\ref{tor2}) and (\ref{tor2-link}), we
have the table of the Alexander polynomials for torus knots
$T(s,2)\,$ and links $L(s,2)\,,$ where  $s=2m+1$ is an integer,
while $m\,$ is an integer or half-integer and equals the degree of
the Alexander polynomial: \be \label{tor2-all} \ba{l}
%{\wt A}_{1}^{2}(q)
{\wt A}_{0}^{1,2}(t)
\equiv A_{0}^{1,2}(t)=1\,,
\vspace{2mm}\\
{\wt A}_{1}^{2,2}(t)
\equiv A_{1\over2}^{2,2}(t)=t^{1\over2}-t^{-{1\over2}},
\vspace{2mm}\\
{\wt A}_{2}^{3,2}(t)\equiv A_{1}^{3,2}(t)=t-1+t^{-1}\,,
\vspace{2mm}\\
{\wt A}_{3}^{4,2}(t)\equiv A_{3\over2}^{4,2}(t)=t^{3\over2}-t^{1\over2}+
t^{-{1\over2}}-t^{-{3\over2}}\,,
\vspace{2mm}\\
{\wt A}_{4}^{5,2}(t)\equiv A_{2}^{5,2}(t)=t^{2}-t+1-t^{-1}+t^{-2}\,,
\vspace{2mm}\\
{\wt A}_{5}^{6,2}(t)\equiv A_{5\over2}^{6,2}(t)=
\vspace{2mm}\\
\quad\quad\quad\,\, =t^{5\over2}-t^{3\over2}+t^{1\over2}-t^{-{1\over2}}+
t^{-{3\over2}}-t^{-{5\over2}}\,,
\vspace{2mm}\\
{\wt A}_{6}^{7,2}(t)\equiv A_{3}^{7,2}(t)=
\vspace{2mm}\\
\quad\quad\quad\,\, =t^{3}-t^{2}+t-1+t^{-1}-t^{-2}+t^{-3}\,,
\vspace{2mm}\\
%{\wt A}_{7}^{8,2}(t)\equiv A_{7\over2}^{8,2}(t)=t^{7\over2}-
%t^{5\over2}+t^{3\over2}-t^{1\over2}+
%t^{-{1\over2}}-t^{-{3\over2}}+t^{-{5\over2}}-t^{-{7\over2}}\,,
%\vspace{2mm}\\
%{\wt A}_{8}^{9,2}(t)\equiv A_{4}^{9,2}(t)=t^{4}-t^{3}+t^{2}-t+1-t^{-1}+
%t^{-2}-t^{-3}+t^{-4}\,,
%\vspace{2mm}\\
\cdots\cdots\cdots\cdots\cdots\cdots\cdots\cdots\cdots
\cdots\cdots\cdots\cdots\cdots
%\cdots\cdots\cdots\cdots\cdots\cdots
\vspace{2mm}\\
{\wt A}_{2m}^{2m+1,2}(t)\equiv A_{m}^{2m+1,2}(t)=
\vspace{2mm}\\
=t^{m}-t^{m-1}+t^{m-2}-\cdots
t^{-m}=\sum\limits_{i=0}^{2m}(-1)^{i}t^{m-i}\,. \ea \ee
  Two different notations for the Alexander polynomials
 are related as
 \[ {\wt
A}_{2m}^{2m+1,2}(t)= A_{m}^{2m+1,2}(t)\,.\]
% denote the
%Alexander polynomials for torus knots and links of type $(s,2)\,,$
%where $s$ now is positive integer, and $m$ is in (\ref{max}).
 Recurrence
relation for the polynomials (\ref{tor2-all}) is given by
(\ref{alex-rec}), which immediately follows from the skein relation
(\ref{alex-skein}), in view of the correspondence (\ref{cor}).

Let us  show that (\ref{alex-rec}) can  be  obtained from
(\ref{alex-rec2}) as well.
%Let us find the connection between the
%coefficients of (\ref{alex-rec}) and (\ref{alex-rec2}).
To see that, we first write  (\ref{alex-skein}) in general form
\be\label{alex-skein-gen} A_{+}(t)=b_{1}A_{O}(t)+b_{2}A_{-}(t)\,.
 \ee
With the account of (3), we also have the recursion
%\be\label{alex-skein}
%A_{+}(t)=(t^{1\over2}-t^{-{1\over2}})A_{O}(t)+A_{-}(t)\,,
% \ee
\be\label{alex-rec-gen} {\wt A}_{n+1}^{2}(t)=b_{1}{\wt
A}_{n}^{2}(t)+b_{2}{\wt A}_{n-1}^{2}(t)\,. \ee
 Then  rewrite eq. (\ref{alex-rec2}) in terms of tilded
 ${\wt A}_{n}^{2}(t)$:
 %(\ref{alex-rec})
 \be\label{alex-rec2-ge} {\wt A}_{n+1}^{2}(t)=(t+t^{-1}){\wt
A}_{n-1}^{2}(t)-{\wt A}_{n-3}^{2}(t)\,. \ee
 The latter in general terms looks as
%(\ref{alex-rec2-ge}) in general case
\be\label{alex-rec2-gen}  {\wt A}_{n+1}^{2}(t)=c_{1}{\wt
A}_{n-1}^{2}(t)+c_{2}{\wt A}_{n-3}^{2}(t)\,. \ee From
(\ref{alex-rec-gen}) we have \be\label{alex-rec-gen2} {\wt
A}_{n}^{2}(t)=b_{1}{\wt A}_{n-1}^{2}(t)+b_{2}{\wt A}_{n-2}^{2}(t)\,,
\ee \be\label{all-rec-gen3} {\wt A}_{n-1}^{2}(t)=b_{1}{\wt
A}_{n-2}^{2}(t)+b_{2}{\wt A}_{n-3}^{2}(t)\,. \ee
 Insert (\ref{alex-rec-gen2}) into (\ref{alex-rec-gen})
  \be\label{all-rec-} {\wt A}_{n+1}^{2}(t)=(b_{1}^{2}+b_{2}){\wt
A}_{n-1}^{2}(t)+b_{1}b_{2}{\wt A}_{n-2}^{2}(t)\,.
\ee
 Then put ${\wt
A}_{n-2}^{2}(t)$ from (\ref{all-rec-gen3})  into (\ref{all-rec-})
\be\label{all-rec--} {\wt A}_{n+1}^{2}(t)=(b_{1}^{2}+2b_{2}){\wt
A}_{n-1}^{2}(t)-b_{2}^{2}{\wt A}_{n-3}^{2}(t)\,. \ee
 Comparison of
(\ref{all-rec--}) and (\ref{alex-rec2-gen}) gives
\be\label{cof}c_{1}=b_{1}^{2}+2b_{2}\,,\quad c_{2}=-b_{2}^{2}\,.\ee
From eq. (\ref{cof}),
 \be\label{cof2} b_{1}=(c_{1}-2b_{2})^{1\over2}\,,\quad
  b_{2}=(-c_{2})^{1\over 2}\,.
 \ee
 Note that the latter two formulas, which involve
 general coefficients, will be used below (see Sec. 4).
 Comparing (\ref{alex-rec2-ge}) and (\ref{alex-rec2-gen}) yields
\[ c_{1}=t+t^{-1}\,,\quad c_{2}=-1\,.\]
 With account of (\ref{cof2}) this implies
\[ b_{2}=1\,,\quad b_{1}=t^{1\over2}-t^{-{1\over2}}\,,\]
which coincides with the coefficients  in (\ref{alex-rec}), and thus
our statement is proved.

Since the formulas (\ref{cof}) and (\ref{cof2}) connect arbitrary pairs $b_1,b_2$
and $c_1,c_2$, this allows one to gain general skein relation from
the corresponding recurrence relation
for the set of torus knots $T(s,2)$.
%It is easy to see that (\ref{alex-rec}) can be also obtained from (\ref{alex-rec2}).
 %Knowing the coefficients $c_{1}\,,\,c_{2}$ for this relation, we
%immediately obtain the corresponding skein relation in the form
%\be\label{skein} A_{+}=b_{1}A_{O}+b_{2}A_{-}\,, \ee where
%$b_{1}\,,\,b_{2}$ are given by (\ref{cof2}) and $A$ denotes some
%polynomial invariant.

\section{Alexander polynomials from Chebyshev polynomials}

In this section we describe the connection between Alexander
polynomials and Chebyshev polynomials, using the $q$-numbers. The
$q$-number corresponding to the integer $n$ is defined as (see
e.g.~\cite{Bi, Ma} and~\cite{Kac})
 \be \label{q-def} [n]_{q}={{\textstyle
{q^{n}-q^{-n}}}\over{\textstyle {q-q^{-1}}}}\,, \ee where $q$ is a
parameter.  If $q\rightarrow 1\,,$ then $[n]_{q}\rightarrow n\,.$
Considering $\,q\,$ to be  variable, we go over to the
$q$-polynomials.
 Some of the $q$-numbers (or $q$-polynomials) are as follows:
% \label{q-ex}\ba{l}
\[ [1]_{q}=1\,,\quad  [2]_{q}=q+q^{-1}\,,
\]
 \[ [3]_{q}=q^{2}+1+q^{-2}\,, \quad  [4]_{q}=q^{3}+q+q^{-1}+q^{-3}\,, %\cdots\cdots\,
 \]
\vspace{-5mm}
 \[
\cdots\cdots\cdots\cdots\cdots\cdots\cdots\cdots\cdots\cdots\cdots\cdots\,
\]
\vspace{-6mm}
 \be\label{q-ex}
 [n]_{q}{=}q^{n-1}+q^{n-3}+\cdots+q^{-(n-1)}=\sum\limits_{i=0}^{n-1}q^{n-1-2i}\,.
 \ee
 The (easily verifiable) recurrence relation
 for the $q$-numbers %(\ref{q-ex})
 (or $q$-polynomials) is
 \be\label{q-rec}
[n+1]_{q}=(q+q^{-1})[n]_{q}-[n-1]_{q}\,.\ee
%\end{multicols}
%\end{document}
From now on, rename the variable in the Alexander polynomials:
\[ t\rightarrow q\,.\]
From comparison, for the considered class of torus knots, of the
Alexander polynomials and the $q$-polynomials, we see that the
Alexander polynomials (\ref{tor2}) can be expressed through
$q$-polynomials (\ref{q-ex})
 in the simple way:
\be\label{om} A_{n}^{2}(q)=[n+1]_{q}-[n]_{q}\,. \ee
 This relation %(\ref{om})
 for the Alexander polynomials
  was found in~\cite{Ga1,Ga2}, in the context of their
  correspondence to masses of vector quarkonia.

Below we will need some properties of the classical Chebyshev
polynomials, in order to formulate the Alexander polynomials in
terms of the Chebyshev ones. If $x=2\cos\theta\,,$ the Chebyshev
polynomials of the first kind are defined as \be\label{T}
T_{n}(x)=2\cos (n\theta)\,.\ee From (\ref{T}), some first cases read
%\be\label{Tx} \ba{l}
\[
T_{0}{=}2\,,\quad
%\vspace{2mm}\\
T_{1}{=}x\,,\quad
% \vspace{2mm}  \\
T_{2}{=}x^{2}-2\,,\quad
% \vspace{2mm}  \\
T_{3}{=}x^{3}-3x\,,\quad \ldots\,\,.\]
%\vspace{2mm}  \\
%T_{4}=x^{4}-4x^{2}+2\,,
%\vspace{2mm}  \\
%T_{5}=x^{5}-5x^{3}+5x\,.
% \ea \ee
The recurrence formula is known as
 \be\label{rec-T}
T_{n+1}=xT_{n}-T_{n-1}\,. \ee Chebyshev polynomials of the second
kind are \be\label{V} V_{n}(x)={\sin((n+1)\theta)\over
{\sin\theta}}\,.\ee Polynomials $\,T_{n}\,$ and $\,V_{n}\,$ are both
monic and have the degree $n$. From (\ref{V}) we have
%\be\label{Vx} \ba{l}
\[
V_{0}{=}1\,,\quad
%\vspace{2mm}\\
V_{1}{=}x\,,\quad
% \vspace{2mm}  \\
V_{2}=x^{2}-1\,,\quad
% \vspace{2mm}  \\
V_{3}=x^{3}-2x\,,\quad\ldots\,,\]
%\vspace{2mm}  \\
%V_{4}=x^{4}-3x^{2}+1\,,
%\vspace{2mm}  \\
%V_{5}=x^{5}-4x^{3}+3x\,,
% \ea \ee
and the recurrence relation is \be\label{rec-Vx}
V_{n+1}=xV_{n}-V_{n-1}\,.\ee
 There is a connection between (\ref{T})
and (\ref{V}) \be\label{con-TV} T_{n}(x)=V_{n}(x)-V_{n-2}(x)\,.\ee

Putting
\be\label{qe}q=e^{i\theta}\ee
 into (\ref{q-def}), we have
\be \label{q-def2} [n]_{q}={{\sin
(n\theta)}\over{\sin\theta}}=V_{n-1}(x)\,, \ee where $\,V_{n}(x)\,$
is the Chebyshev polynomial of the second kind, and
\be\label{xq}x=2\cos\theta=q+q^{-1}\,.\ee From (\ref{q-def2}),
(\ref{xq}) it is seen that \be\label{vnq}V_{n}(q)=[n+1]_{q}\,,\ee
and therefore (\ref{om}) takes the form \be\label{om2}
A_{n}^{2}(q)=V_{n}(x)-V_{n-1}(x)\,, \hspace{8mm} x=q+q^{-1} . \ee
 %,\quad    x=q+q^{-1}\,.\ee
%recurrence relation (\ref{q-rec}) looks
%\be\label{q-rec2} [n+1]_{q}=xV_{n-1}(x)-V_{n-2}(x)\,.\ee
Thus, the Alexander polynomials $A_{n}^{2}(q)$ are obtained from the
Chebyshev polynomials of the second kind $V_{n}(x)$ (\ref{V}), after
changing the variables $x\rightarrow q+q^{-1}\,,$  by means of the
formula (\ref{om2}).

\section{Generalized Alexander polynomials and HOMFLY polynomials}

Now let us put into consideration the $q,\!p$\,-numbers, a natural
generalization of $q$-numbers. With the help of $q,p$-numbers we
will construct a generalization of the Alexander polynomials --
$A_{n}^{2}(q,p)\,,$ which now depend on two variables $\,q,p\,.$
Afterwards we intend to show that $\,A_{n}^{2}(q,p)\,$ turn into
the well-known HOMFLY polynomials by an appropriate change of
variables.

The $q,p$-number corresponding to the integer number $n$ is defined
as (see e.g.~\cite{CJ})
 \be \label{qp-def}
[n]_{q,p}={{\textstyle {q^{n}-p^{n}}}\over{\textstyle {q-p}}}\,, \ee
where $q\,,p$ are some complex parameters. If $p=q^{-1}\,,$ then
$[n]_{q,p}=[n]_{q}\,.$
 Some of the $q,p$-numbers are
%\be\ba{l}
 \[ [1]_{q,p}=1\,,\quad [2]_{q,p}=q+p\,,
  \]
 \[ [3]_{q,p}{=}q^{2}+qp+p^{2}\,,\quad   [4]_{q,p}{=}q^{3}+q^{2}p+qp^{2}+p^{3}\,, \]
 \vspace{-2mm}
 \[ \cdots\cdots\cdots\cdots\cdots\cdots\cdots\cdots\cdots\cdots\cdots\,
 \]
\vspace{-2mm}
 \[ [n]_{q,p}=q^{n-1}+q^{n-2}p+q^{n-3}p^{2}+\cdots+\]
 \vspace{-5mm}
 \be\label{qp-ex}
+qp^{n-2}+p^{n-1}{=}\sum\limits_{i=0}^{n-1}q^{n-1-i}p^{i}=q^{n-1}
\sum\limits_{i=0}^{n-1}q^{-i}p^{i}\,. \ee
 Considering $q$ and $p$ as variables, we deal with $q,p$-polynomials.
  Then, the recurrence relation for them
is  \be\label{qp-rec}
[n+1]_{q,p}=(q+p)[n]_{q,p}-qp[n-1]_{q,p}\,.\ee

On the base of eq. (\ref{vnq})
%and (\ref{om2}), using
and the expression (\ref{qp-def}) or (\ref{qp-ex}) for the
$q,p$-polynomials , we introduce a natural generalization of the
Chebyshev polynomials of the second kind, which now depend on the
two variables:
 \be\label{cheb2} V_{n}(q,p)=[n+1]_{q,p}\,.\ee
  %As it follows
  From (\ref{qp-rec}), (\ref{cheb2}) the recurrence
relation does follow: \be\label{vqp-rec}
V_{n+1}(q,p)=(q+p)V_{n}(q,p)-qpV_{n-1}(q,p)\,.\ee
 Now, in analogy with (\ref{om2}), we introduce the two-variable
 generalized Alexander
polynomial  as linear combination of the polynomials (\ref{cheb2}).
 Due to this proposal, the following recurrence formula takes place:
\be\label{aqp-rec}
A_{n+1}^{2}(q,p)=(q+p)A_{n}^{2}(q,p)-qpA_{n-1}^{2}(q,p)\,.\ee
 This is a direct analog of (\ref{alex-rec2}) and reduces to it if $q=t$ and
$p=t^{-1}\,.$ To continue the analogy we take \be\label{2first}
A_{0}^{2}(q,p)=1\,,\,\, \ \ \ A_{1}^{2}(q,p)=q-qp+p\,.\ee
 It is easy to see that (\ref{aqp-rec}) with (\ref{2first}) will be valid if
\be\label{om2gen}\ba{l} A_{n}^{2}(q,p)=V_{n}(q,p)-qpV_{n-1}(q,p)=
\vspace{2mm}  \\
=[n+1]_{q,p}-qp[n]_{q,p}\,. \ea\ee

Setting \be\label{qpe}q=re^{i\theta}\,,\quad p={\bar
q}=re^{-i\theta}\,\ee into (\ref{qp-def}), we have \be
\label{qp-def2} [n]_{q,p}={{r^{n}\sin
(n\theta)}\over{r\sin\theta}}=r^{n-1}V_{n-1}(x)\,. \ee If $\,r=1\,,$
eq. (\ref{qp-def2}) turns into (\ref{q-def2}).
 Taking into account (\ref{cheb2}) and (\ref{qp-def2}), we obtain
 $V_{n}(q,p)$ with a factorized form of dependence on the variables $r,x:$
\be\label{vrx}V_{n}(r,x)=r^{n}V_{n}(x)\,.\ee Here $V_{n}(x)$ is the
classical Chebyshev polynomial of second kind, with $x$ as in
(\ref{xq}). The corresponding two-variable Chebyshev polynomials of
 first kind also arise:
% \be\label{T2}
\[ T_{n}(r,x)=2r^{n}\cos (n\theta)\,.\]
%\ee
In the variables $r,x\,,$  see (\ref{qpe}) and (\ref{xq}), the
recurrence relation (\ref{aqp-rec}) is written  as
\be\label{arx-rec}
A_{n+1}^{2}(r,x)=rxA_{n}^{2}(r,x)-r^{2}A_{n-1}^{2}(r,x)\,.\ee
 The first two polynomials (\ref{2first}) become \be\label{2first-rx}
A_{0}^{2}(r,x)=1\,,\quad A_{1}^{2}(r,x)=rx-r^{2}\,.\ee
 From (\ref{om2gen}) and (\ref{vrx}) we also have
\be\label{om3gen}
A_{n}^{2}(r,x)=r^{n}(V_{n}(x)-rV_{n-1}(x))\,.
\ee

Now we make a key proposal: we apply the generalized Alexander
polynomials $A_{n}^{2}(r,x)\,,$ given by (\ref{arx-rec}) and
(\ref{2first-rx}), for describing the torus knots $T(s,2)\,.$
 From (\ref{cof2}), with account of (\ref{qpe}), we have
\[c_{1}=rx\,,\quad c_{2}=-r^{2}\,,\]
and then \be\label{cof-skein} b_{2}=r\,,\quad b_{1}=(rx-2r)^{1\over
2}=r^{1\over 2}(x-2)^{1\over 2}\,.\ee
 Hence, as a generalization of (\ref{alex-skein}), from (\ref{alex-skein-gen}),
  (\ref{alex-rec-gen}) %(\ref{skein})
 and (\ref{cof-skein}), we obtain
 the skein relation for the generalized
Alexander polynomials: \be\label{skein-arx} A_{+}(r,x)=r^{1\over
2}(x-2)^{1\over 2}A_{O}(r,x)+rA_{-}(r,x)\,. \ee

Now let us explore the connection between the generalized Alexander
skein relation (\ref{skein-arx})
 and the HOMFLY skein relation.
By definition, the HOMFLY polynomials $H(a,z)$ satisfy the skein relation
\[ a^{-1}H_{+}(a,z)-a^{1}H_{-}(a,z)=zH_{O}(a,z)\,,\]
or, in  equivalent form, \be\label{hom-skein}
H_{+}(a,z)=azH_{O}(a,z)+a^{2}H_{-}(a,z)\,, \ee with
$H_{unknot}=1\,.$
 As before, consider the torus knots $T(s,2)\,,$ $s$
odd integer. For these, the notation for the corresponding HOMFLY
polynomials is similar to the notation for the Alexander ones,
namely
$$H(s,2)(a,z)\equiv H(2m+1,2)(a,z)\equiv H_{m}^{2}(a,z)\equiv H_{m}^{2}\,.$$
The short list of the HOMFLY polynomials for torus knots
$\,T(s,2)\equiv T(2m+1,2)$ is:
\vspace{-5mm}
 \be \label{tor2h} \ba{l}
H_{0}^{1,2}(a,z)=1\,,
\vspace{2mm}\\
%H(3,2)(a,z)\equiv
H_{1}^{3,2}(a,z)=2a^{2}+a^{2}z^{2}-a^{4}\,,
\vspace{2mm}\\
%H(5,2)(a,z)\equiv
H_{2}^{5,2}(a,z)
%\vspace{2mm}\\
=3a^{4}+4a^{4}z^{2}+a^{4}z^{4}-2a^{6}-a^{6}z^{2}\,,   %\,\cdots\,.\ea \ee
\vspace{2mm}\\
%H(7,2)(a,z)\equiv
 H_{3}^{7,2}(a,z)
%\vspace{2mm}\\
=4a^{6}+10a^{6}z^{2}+6a^{6}z^{4}+a^{6}z^{6}-3a^{8}-
\vspace{2mm}\\
%\vspace{2mm}\\\qquad\qquad\qquad\qquad\qquad\,\,\,{}-
\qquad\quad\qquad -4a^{8}z^{2}-a^{8}z^{4}\,,\\
%\,\cdots\cdots\,.
\vspace{-2mm}
\cdots\cdots\cdots\cdots\cdots\cdots\cdots\cdots\cdots\cdots\cdots\cdots\cdots\cdots\,.
\ea\ee
Recurrence relation for (\ref{tor2h}) reads  %can be proven by induction
\be\label{hom-rec2}
H_{m+1}^{2}(a,z)=a^{2}(z^{2}+2)H_{m}^{2}(a,z)-a^{4}H_{m-1}^{2}(a,z)\,.
\ee

If we compare (\ref{hom-rec2}) and (\ref{arx-rec}), we see that
through the substitution
 \be\label{ha} r=a^{2}\,,\quad
x=z^{2}+2\,,\ee
 the HOMFLY polynomials and the generalized Alexander
polynomials coincide:
\[ H_{n}^{2}(a,z)=A_{n}^{2}(r,x)=r^{n}(V_{n}(x)-rV_{n-1}(x))\,.\]
Then, the HOMFLY skein relation (\ref{hom-skein}) in the variables
$r,x$, see (\ref{ha}), looks as
%\be\label{hom-skein}
%$$r^{-{1\over2}}H_{+}(r,x)-r^{1\over2}H_{-}(r,x)=(x-2)^{1\over2}H_{O}(r,x)\,.$$
\be\label{skein-hrx}
H_{+}(r,x)=r^{1\over 2}(x-2)^{1\over 2}H_{O}(r,x)+rH_{-}(r,x)\,,
\ee
which coincides with (\ref{skein-arx}).
Besides,
%\be\label{ah}
\[ A_{0}^{2}(r,x)=H_{0}^{2}(r,x)\,,\quad A_{1}^{2}(r,x)=H_{1}^{2}(r,x)\,.\]

Thus, we have proved that the generalized Alexander polynomials and
their skein relation go over into the HOMFLY ones by applying the
parametrization (\ref{ha}). On the other hand, the HOMFLY skein
relation and polynomials turn into the generalized Alexander ones
with the help of inverse substitution
\[ a=r^{1\over 2}\,,\quad z=(x-2)^{1\over 2}\,.\]

\section{Concluding remarks}

We have demonstrated that the connection of the Chebyshev
polynomials with the Alexander polynomials can be realized in a
rather simple way if one uses, as an auxiliary tools, the concept of
$q$-numbers. On the other hand, existence of the $q,\!p$\,-numbers,
which generalize the $q$-numbers, makes it possible to generalize
the Chebyshev polynomials to their two-variable modification and, by
exploiting the analogy with previous one-variable case, also to
achieve two-variable generalization of the Alexander polynomials.
Finally, we have found that the two-variable extended Alexander
polynomials are mapped onto the HOMFLY polynomials.

We hope the proposed way of usage of the Chebyshev polynomials will
be helpful for further investigation of knots and links, not only on
the base of the Alexander polynomials (along with their two-variable
modification) and HOMFLY polynomials treated above, but also
possibly in connection with Kauffman polynomials and other known
polynomial invariants.
 Besides, it is of interest to study, within the proposed scheme,
 the $(s,r)$ torus knots more general than the particular class $(s,2)$
considered in this paper. Subsequently we hope to use the explored
polynomial invariants within the framework of some physical models.

\bigskip
%{\bf Acknowledgements}
 This research was partially supported by the
Grant 29.1/028 of the State Foundation of Fundamental Research of
Ukraine, and by the Special Program of the Division of Physics and
Astronomy of the NAS of Ukraine.

\begin{flushleft}

\end{flushleft}

\rezume{%
ПРО ПОЛІНОМИ ЧЕБИШОВА І ТОРИЧНІ ВУЗЛИ}
%СФЕРИЧНА МОДЕЛЬ ДЛЯ КІЛЬКІСНОГО\\ АНАЛІЗУ~ НАНЕСЕНИХ~
%КАТАЛІЗАТОРІВ\\ МЕТОДОМ РФС: СИСТЕМИ З НИЗЬКОЮ\\ ПИТОМОЮ
%ПОВЕРХНЕЮ}
{%
О.М. Гаврилик, А.М. Павлюк} {В роботі показано, що $q$-числа та їх
двопараметричні узагальнення, $q,p$-числа, можна використати для
отримання деяких поліноміальних інваріантів торичних вузлів і
зачеплень.
 По-перше, показується, що $q$-числа, які тісно пов'язані з поліномами Чебишова,
 можуть бути пов'язані з поліномами Александера для класу  $T(s,2)\,$
торичних вузлів, де $s\,\,$ - непарне ціле число, і використані для
знаходження відповідного скейн-співвідношення. Затим, ми
застосовуємо цю процедуру для отримання, за допомогою $q,p$-чисел,
двопараметричних узагальнених поліномів Александера, та показуємо
зв'язок останніх із поліноміальними інваріантами HOMFLY та їх
скейн-співвідношенням.}

\end{multicols}
\end{document}